\documentclass[prl, twocolumn, showpacs, amsmath]{revtex4}
\usepackage{dcolumn}
\usepackage{bm}
\usepackage{graphicx}
\usepackage{color}
\DeclareGraphicsExtensions{. jpg,. pdf, . mps, . png, . eps, . ps, . EPS}

\begin{document}
\def\be{\begin{equation}}
\def\ee{\end{equation}}

\def\bc{\begin{center}} 
\def\ec{\end{center}}
\def\bea{\begin{eqnarray}}
\def\eea{\end{eqnarray}}
\newcommand{\avg}[1]{\langle{#1}\rangle}
\newcommand{\Avg}[1]{\left\langle{#1}\right\rangle}

\title{Critical fluctuations in spatial complex networks}

\author{Serena Bradde$^1$, Fabio Caccioli$^1$, Luca Dall'Asta$^2$ and Ginestra Bianconi$^3$}

\affiliation{$^1$ International School for Advanced Studies, via Beirut
 2/4, 34014, Trieste, Italy\\
$^2$ The Abdus Salam International Center for Theoretical 
Physics, Strada Costiera 11, 34014 Trieste, Italy \\
$^3$ Department of Physics, Northeastern University, Boston, 
Massachusetts 02115 USA}
\begin{abstract}An anomalous mean-field solution is known to capture the non
 trivial phase diagram of the Ising model in annealed complex
 networks. Nevertheless the critical fluctuations in random complex
 networks remain mean-field. 
 Here we show that a break-down of this scenario can be obtained when complex networks are embedded in geometrical spaces.
Through the analysis of the Ising model on annealed spatial networks, we reveal in particular the spectral properties of networks responsible for critical fluctuations and we generalize the Ginsburg criterion to complex topologies.
\end{abstract}

\maketitle

Large attention  has been recently addressed to the effects that different topological properties may induce on the behavior of equilibrium and non-equilibrium processes defined on networks and to the possible implications for the study of several social, biological and technological networks \cite{crit,Dynamics}. 
Heterogeneous degree distributions, small world and spectral properties, in particular, have been recognized as responsible of novel types of phase transitions and universality classes \cite{crit,Dynamics,Vespignani,Synchr}.
For instance, scale-free networks present a complex critical behavior for the Ising model, percolation and spreading processes, that 
explicitly depends on the exponent of the power-law in the degree distributions  \cite{crit,Dynamics,Vespignani}. On the other hand, the existence of non trivial spectral properties is crucial for  the stability of synchronization processes and $O(n)$ models \cite{Synchr}.

Despite the large interest in the subject, much smaller attention has been devoted  to critical phenomena on complex networks embedded in a metric space \cite{Alain, Vespignani2,Manna,Newman1,Boguna}, though
some important problems related to navigability, efficiency and search optimization in spatial networks have been already discussed in the literature \cite{Kleinberg,Latora,Newman2,Havlin2}. In fact, spatial embedding is a very relevant aspect of infrastructure  and technological networks, including airports networks, the Internet, and power-grid networks. Moreover, a pivotal role in shaping the topology of social networks is played by  hidden metric structures in some underlying abstract space, such as that of the social distance between individuals \cite{Newman1,Boguna}.

The aim of this Letter is to investigate the role of spatial embedding in relation with the critical behavior of phase transitions in complex networks. It is well known that in regular lattices, space dimensionality governs  the critical behavior of equilibrium and non-equilibrium systems. In particular, below the upper critical dimension, critical fluctuations that are not  captured by the mean-field approach set in. Similarly, for complex networks embedded in a low dimensional space we can expect that, as the link probability becomes short ranged, the effect of the underlying space might change the critical behavior leading to a break-down of the validity of (heterogeneous) mean-field arguments. This should be relevant to understand  real phenomena  in spatial networks, such as the spreading of viruses \cite{Vespignani2}, the emergence of congested phases in the packet-based traffic on technological networks \cite{Moreno} and cascading failure phenomena in powe
 r-grid networks \cite{Motter3}.

As a prototypical example of the complex behavior induced by spatial embedding, in this Letter we consider the Ising model on annealed scale-free networks. 
On a scale-free  network  with a degree distribution $P(k)\sim k^{-\gamma_{SF}}$, the critical temperature of the Ising model diverges for $\gamma_{SF}<3$. The critical exponents, computed  by means of the annealed network approximation \cite{ising_g} or by assuming a quenched randomness \cite{Doro_ising,Doro2}, deviate from the mean-field ones as long as $\gamma_{SF}<5$, with the exception of $\gamma,\gamma'$  describing the divergence of the magnetic susceptibility  $\chi$ close to the critical temperature $T_c$, ( $\chi\sim |T-T_c|^{-\gamma,\gamma'}$). In fact $\gamma,\gamma'$ remain always fixed to their mean field value $\gamma=\gamma'=1$. For these reasons we refer to the critical behavior of random scale-free networks  as the {\em heterogeneous mean-field} solution. 
We derive here a {\em Ginsburg criterion} \cite{Lebellac} for spatial complex networks determining the condition under which critical fluctuations become larger than the ones predicted within a mean-field approach. In particular, we will show that the critical behavior is always mean-field, whenever the matrix ${\bf p}= \{p_{ij}\}_{i,j = 1,\dots,N}$, fixing the probabilities of existence of each link $(i,j)$  has a finite spectral gap $\Delta$ between the maximal eigenvalue $\Lambda$ and the second maximal one $\lambda_2$. On the contrary, when the spectral gap $\Delta\to 0$ in the thermodynamic limit, the critical behavior depends on the behavior of the tail of the spectrum of  $p$. We will demonstrate by theoretical and numerical results that the behavior of such tail is well captured by an exponent $\delta_S$, related to the effective dimension $d_{eff}$ of the network through the relation $\delta_S=(d_{eff}-2)/2$. We find that for $\delta_S<1$ the critical fluctuations be
 come dominant and close enough to the critical temperature the mean-field theory is not sufficient to correctly characterize the critical exponents, possibly calling for renormalization group calculations.

{\it Networks with spatial embedding -}
We consider networks of $N$ nodes embedded in
a $d$-dimensional euclidean metric space, each node $i=1,\ldots,N$ having position $\vec{r}_i$. 
The minimal hypothesis \cite{entropy} that can be made on random networks with 
heterogeneous degrees and spatial embedding is that links $(i,j)$ are
drawn with probability $p_{ij}$ given by 
\be
p_{ij}=\frac{\theta_i \theta_j J(\vec{r}_i,\vec{r}_j)}{1+\theta_i
 \theta_j J(\vec{r}_i,\vec{r}_j)}\simeq\theta_i\theta_j J(\vec{r}_i,\vec{r}_j)
 \label{pij2}
\ee
where we assumed that $[\max_i(\theta)]^2[\max_{\vec{r},\vec{r}'}
J(\vec{r},\vec{r}')]\ll 1$ and that the matrix $J(\vec{r}_i,\vec{r}_j)$ only depends on the distance
between the nodes, i.e. $
J(\vec{r}_i,\vec{r}_j)=J(|\vec{r}_i-\vec{r}_j|). $
In this ensemble the degree $k_i$ of a node $i$ is a Poisson random variable with expected degree $\overline{k_i}$ fixed by means of the hidden variables $\theta_i$ and given by the
relation $\overline{k_i}=\sum_j p_{ij} $. 
Therefore, given a set of expected degrees $\{\overline{k_i}\}$, we can evaluate the $\{\theta_i\}$ variables by solving the equations $\overline{k}_i=\sum_jp_{ij}$. 
Networks with homogeneous degrees are generated by fixing  
$\theta_i=\theta \ \forall i$, that corresponds to the Manna-Sen model of spatial networks \cite{Manna}. 
Another special choice is that of space-independent couplings $J_{ij}=J$ $\forall i,j = 1,\dots,N$, that gives
$\theta_i= {\overline{k}_i}/{\sqrt{J\avg{\overline{k}}N}}$ where $\langle \overline {k}\rangle=\sum_i \overline {k}_i/N$. In this case, our formalism easily 
recovers known results for both the percolation threshold and the critical temperature of the Ising model on complex networks without spatial embedding.

{\it Ising model on annealed spatial networks and the Ginsburg criterion -}
We consider a system of binary spin variables 
$s_i=\pm 1$, for $i=1,\dots,N$, defined on the nodes of a given annealed network with spatial embedding and link probability
given by the matrix ${\bf p}$. 
The partition function \cite{ising_g,crit} for this problem is given by 
\be
Z=\sum_{\{{s}_i\}} e^{-\beta H(\{{s}_i\})}
\ee
with
\be
H({s}_i)=-\frac{1}{2}\sum_{i\neq j}s_i\theta_iJ_{ij}\theta_j{s}_j
-\sum_i {H}_i {s}_i. 
\ee
In order to derive a Ginsburg criterion for this statistical mechanics
problem, we generalized the classical approach by means of stationary
phase approximation \cite{Lebellac}. Considering only the first-order terms in the expansion leads to mean-field results. Thus the validity of the mean-field solution can be checked by evaluating the higher order corrections at the critical point.
Critical fluctuations that are neglected by the mean-field set in when  the second order corrections diverge dominating the behavior of the  susceptibility at the criticality. 
In the stationary phase approximation, the magnetization of the system is given by the $m_i^{0}$'s
satisfying the self consistent equations
\begin{equation}
m_i^0=\tanh[\beta(H_i+\sum_j \theta_i J_{ij}\theta_j m_j^0)]. 
\end{equation}
At the second order of the stationary phase approximation \cite{Lebellac}, performing a Legendre transformation we can evaluate the free energy $\Gamma(\{m_i\})$ as
\bea
&\Gamma&(\{m_i\}) =-\frac{1}{2}\sum_{ij} m_i\theta_iJ_{ij}\theta_j
m_j\nonumber \\ &+& \frac{1}{2\beta}\sum_i
[(1-m_i)\ln(1-m_i)+(1+m_i)\ln(1+m_i)]\nonumber \\ &+&\frac{1}{2z\beta} \ln \det
[\delta_{ij}-\beta J_{ij}\theta_i\theta_j(1-m_j^{2})]
\eea 
where  the external field $H_i=\partial \Gamma(\{m_i\})/\partial m_i$ and we have introduced the parameter $z$ in order to keep track of the different orders in the expansion. 
The susceptibility matrix is defined as $\chi_{i,j}^{-1}=\frac{\partial^2\Gamma}{\partial m_i\partial m_j}$. 
We compute it in the paramagnetic phase, where $m_i=0$, and then we perform the projection along the eigenvector $u_i^{\lambda}$ associated to the eigenvalue $\lambda$ of the connectivity matrix, obtaining
\bea
\chi^{-1}_{\lambda}&=&-\lambda+\frac{1}{\beta}+\frac{1}{z}\sum_{i ,\ell} p_{i\ell}
\left[{\bf 1}-\beta {\bf p}\right]^{-1}_{\quad\ell i}(u^{\lambda}_i)^2\,,
\eea
where ${\bf 1}$ is the identity matrix.
The instability of the paramagnetic phase is now determined in terms of the largest eigenvalue $\Lambda$ of the matrix $p_{ij}$ through the condition $\chi_{\Lambda}^{-1}(T_c)=0$.
 If we express the susceptibility in terms of the spectral density $\rho(\lambda)$ of the  matrix ${\bf p}$ as
\be
\chi_{\Lambda}^{-1}(T)=-\Lambda+T+\frac{1}{z}\int d\lambda \rho(\lambda)\frac{\lambda}{1-\frac{\lambda}{T}},
\label{chi}
\ee
to leading order in $1/z$  the critical temperature $T_c$ is given by
\be
T_c=\Lambda-\frac{1}{z}\int d\lambda
\rho(\lambda)\frac{\lambda}{1-\frac{\lambda}{\Lambda}}\,. 
\label{Tc}
\ee
Using $(\ref{Tc})$, we can express the susceptibility, Eq. $(\ref{chi})$, for $T \to T_c$ as
\be
\frac{\chi^{-1}}{T-T_c}=\left[1-\frac{1}{z} \int d\lambda \frac{\rho(\lambda)(\lambda)^2}{(T-\lambda)(T_c-\lambda)}\right].
\label{chi1}
\ee

We assume now that the spectrum $\rho(\lambda)$ 
has a spectral edge $\lambda_c$ equal to the average value of the second largest eigenvalue $\lambda_2$ of $p$, i.e. $\lambda_c=\avg{\lambda_2}$ such that the spectrum for $\lambda<\lambda_c$ is self-averaging. For $\lambda <\lambda_c$, close to the upper edge, we assume the scaling behavior
\be
\rho(\lambda)\simeq (\lambda_c-\lambda)^{\delta_S}
\label{scaling}
\ee
that we can use to perform the integral in $(\ref{chi1})$. Moreover we  define the spectral gap $\Delta_N$ of a network of size $N$ as the difference between the maximal eigenvalue $\Lambda$ and the spectral edge, i.e. $\Delta_N=\Lambda-\lambda_c$. 
Performing a straightforward calculation under the assumption that the  gap $\Delta_N$ is self-averaging in the thermodynamic limit,  i.e. $\lim_{N\to \infty}\Delta_N\to \Delta$, we distinguish two possible behaviors.
If $\Delta>0$, close to the critical temperature $T\to T_c$, we have
\begin{equation}
\chi^{-1}=(T-T_c)[1-(\Delta^{\delta_S-1}{\cal C}_2-{\cal C}_1)/z]
\end{equation}
where ${\cal C}_{1, 2}$ are constants. In this case the critical fluctuations are always mean-field. 
On the other hand, if $\Delta = 0$, we have
\begin{equation}
\chi^{-1}=(T-T_c)[1-(T-T_c)^{\delta_S-1}{\cal C}_3/z+{\cal C}_1/z]
\end{equation}
with constants ${\cal C}_{1, 3}$. In this case the critical behavior depends on the particular value of $\delta_S$. 
For $\delta_S\geq 1$ the corrections of order $1/z$ to
$\chi^{-1}$ do not modify the critical behavior of the
susceptibility. On the contrary, for $\delta_S<1$, the corrections of order $1/z$ diverge close to the
phase transition, the fluctuations dominate the critical
behavior and the mean-field approach cannot be applied.\\ 
As a first check we look at the case of homogeneous degree distributions.
We consider a $d$-dimensional lattice of linear  size $L$, homogeneous hidden variables $\theta_i = \theta$  $\forall i$ and coupling matrices 
\bea
J(\vec{r}_i ,\vec{r}_j)&=&\exp(|\vec{r}_i-\vec{r}_j|/d_0) 
\label{exp}
\eea 
depending on the typical distance $d_0$.
In this case , we get always $\lim_{N\to \infty} \Delta_N=0$ and 
$\delta_S=(d-2)/2$ recovering the classical result of the Ginsburg criterion which states that the critical dimension for the Ising model is $d=4$.


{\it Application to complex spatial networks -}
We now turn to the  case of  linking matrices $p$ describing annealed scale-free networks
embedded in a $d$-dimensional  space with finite critical temperature $T_c$. 
For the sake of concreteness, we consider a regular $d=2$ lattice of side $L$ and we  
 assign to each of the $N=L^2$ nodes an expected degree $\overline{k}$ according to a power-law
distribution $p(\overline{k})\sim \overline{k}^{-\gamma_{SF}}$ and we consider exponentially decaying couplings as in Eq. $(\ref{exp})$.
The values of the parameters $\{\theta_i\}$ are given by the solution of the set of equations $\overline{k}_i=\sum_j p_{ij}$ with $p_{ij}=\theta_i\theta_j J_{ij}$. 
\begin{figure}
\includegraphics[width=0.6\columnwidth, height=50mm]{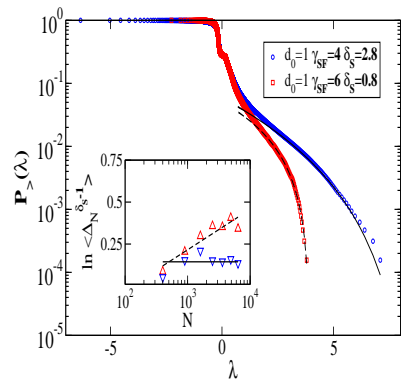}
\caption{(Color online) Cumulative rank plot of the averaged spectra of 100 matrices $p$ for  scale-free random networks embedded in dimension $d=2$, linear size $L=80$, coupling scale $d_0=1$, minimal expected degree $m=2$ and  $\gamma_{SF}=4,6$. 
The behavior for large eigenvalues is well fitted by the expression  $(\ref{scaling})$ with different exponent $\delta_S$ below and above the critical value $\delta_S^{\star}=1$. Inset:  average value $\avg{\Delta_N^{\delta_S-1}}$ for network ensembles with the same parameters as before but with varying system size $N=L^2$. The fit shows that for $\gamma_{SF}=6$ the quantity $\Avg{\Delta_N^{\delta_S-1}}$ increases with the system size, while for $\gamma_{SF}=4$ it remains constant.  }
\label{fig1}
\end{figure}
The role played by spatial embedding in the critical behavior of these  networks is well characterized by the spectrum $\rho(\lambda)$ of the the corresponding  matrix ${\bf p}$. 
For small $d_0$, where we expect non trivial effects of space, the behavior of the spectrum close to $\lambda_c$ follows Eq. $(\ref{scaling})$. 
In Figure $\ref{fig1}$ we report the cumulative distribution (rank plot) of the eigenvalues of $p$ for $d_0=1$ and different values of $\gamma_{SF}$. We observe that the spectral density below the spectral edge is self-averaging and the exponent  $\delta_S$ is a decreasing function of $\gamma_{SF}$ (at  constant $d_0$) assuming values above and below $\delta_S^{\star}=1$. 
However,  the maximal eigenvalue $\Lambda$ and the spectral gap $\Delta$ are not, in general, self-averaging, being subject to strong fluctuations also for large network sizes.
This occurs also for the parameters values studied in Fig. $\ref{fig1}$.  The absence of self-averaging is also observed for networks without spatial embedding, where it is essentially driven by the cutoff fluctuations\cite{finite}. While this \emph{anomalous} effect might be present also in spatial networks, 
it seems that the sample-to-sample fluctuations observed in the spectral 
gap are mainly due to a new feature of spatial networks, i.e.  their local geometry. 
In fact the non self-averaging properties appear also for values of the 
exponent $\gamma_{SF}$ (for example $\gamma_{SF}=6$), where the fourth moment of the degree converges 
and the critical behavior associated to complex network without 
spatial embedding is self-averaging \cite{finite}.
We checked numerically in a number of cases that the spectral gap is non 
self-averaging, but the probability $P(\Delta_N^{\delta_S-1})$ is stable when 
the value of $\Delta_N^{\delta_S-1}$ is rescaled with its average value $\Avg{\Delta^{\delta_S-1}}$ (See Fig. $\ref{fig2}$).
Therefore, in this case we characterize the average critical behavior of the ensemble
by the quantity
\begin{equation}
\Psi=\lim_{N\to \infty}\Avg{\frac{\chi^{-1}}{T-T_c^{eff}}} = \lim_{N\to \infty}\left[1-\frac{\Avg{\Delta_N^{\delta_S-1}}{\cal C}_2-{\cal C}_1}{z}\right]
\end{equation}
where $T_c^{eff}$ is the effective critical temperature of a network and depends explicitly on the size $N$.
If $\Psi$ diverges, i.e. $\lim_{N\to \infty}\Avg{\Delta_N^{\delta_S-1}}\to \infty$,
we expect that the critical fluctuations neglected by the mean-field approach 
become relevant. In the inset of Figure $\ref{fig1}$ we report 
$\avg{\Delta_N^{\delta_S-1}}$ averaged over $100$ realizations of the $p$ matrices for the two network ensembles with $d_0=1$, $\gamma_{SF}=4, 6$ as a function of the network size $N$. 
The results for $d_0=1$, $\gamma_{SF}=4$ are compatible with a limit $\avg{\Delta_N^{\delta_S-1}}\to \mbox{const}$ for $N\to \infty$. Therefore in this case,  the critical behavior should be well captured by the mean-field behavior. For networks with $d_0=1$ and $\gamma_{SF}=6$, instead, $\avg{\Delta_N^{\delta_S-1}}$ seems to diverge as $N\to \infty$, signalling the presence of  critical fluctuations not captured by the mean-field approach. 

\begin{figure}
\begin{center}
\includegraphics[width=.6\columnwidth]{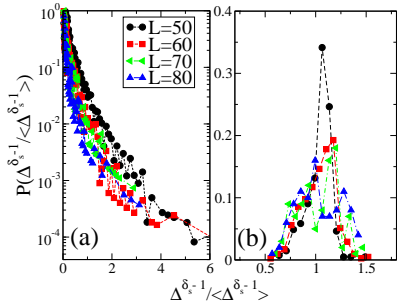}
\end{center}
\caption{(Color online) The distribution $P(\Delta^{\delta_S-1})$ is not self-averaging but is a stable distribution when the variable $\Delta^{\delta_S-1}$ is 
normalized with its average value.
Fixing the value of $d_0=1$, we show in plot (a) the distribution for $\gamma_{SF}=4$, while in (b) $\gamma_{SF}=6$. Both figures are obtained from a diagonalization of $M$ matrices of size $N=L^2$ with $L=50,60,70,80$. In particular the number of samples $M$ is respectively $M=1000$ for $L=50$, $M=400$ for $L=60$, $M=100$ for $L=70$ and $M=100$ for $L=80$. }
\label{fig2}
\end{figure}

{\it Conclusions -}
In this Letter we have investigated how spatial embedding can affect the critical behavior around a phase transition in systems defined on spatial complex networks. In particular, by means of a detailed study of the Ising model on annealed spatial complex networks, we have shown that relevant critical fluctuation not captured by any (heterogeneous) mean-field theory may
set in. Our analysis points out that the knowledge of the spectral properties of the link probability matrix ${\bf p}$ is crucial for the understanding of the critical behavior of dynamical processes and suggests a classification of the latter based on a generalized Ginsburg criterion. More precisely, when the spectrum presents a finite gap $\Delta>0$ in the thermodynamic limit, the fluctuations are always mean-field. If instead the gap vanishes in the thermodynamic limit, the critical behavior depends on the exponent $\delta_S$ describing the scaling of the spectral density close to its upper edge. 
A fascinating open problem is the relation between  the critical behavior of annealed and quenched spatial networks. The solution of this problem might show other  new unexpected effects due to fluctuations of the local geometry.
Finally our results open new perspectives for the comprehension of critical phenomena  in spatial complex networks, whereas the general formalism presented here could be applied to the study of realistic models of 
epidemic spreading in transportation networks as well as of the control of fluctuations in technological and power-grid networks.


\begin{thebibliography}{99}
\bibitem{crit} S. N. Dorogovtsev, A. Goltsev and J. F. F. Mendes, Rev. Mod. Phys. {\bf 80}, 1275 (2008);
\bibitem{Dynamics}
A. Barrat, M. Barth\'elemy, A. Vespignani {\it Dynamical Processes on complex Networks} (Cambridge University Press, Cambridge, 2008). 
\bibitem{Vespignani}
R. Pastor-Satorras and A. Vespignani, Phys. Rev. Lett. {\bf 86}, 3200 (2001); R. Cohen, K. Erez, D. Ben-Avraham, S. Havlin, Phys. Rev. Lett. {\bf 85}, 4626 (2000);
T. Nishikawa, A. E. Motter, Y.-C. Lai and F. C. Hoppensteadt, 
Phys. Rev. Lett. {\bf 91}, 014101 (2003); H. Hong, M. Ha and H. Park, Phys. Rev. Lett. {\bf 98}, 258701 (2007). 
\bibitem{Synchr}
M. Barahona and L. M. Pecora, Phys. Rev. Lett. {\bf 89}, 054101 (2002);
R. Burioni, D. Cassi M. P. Fontana and A. Vulpiani, Europhys. Lett. , {\bf 58}, 806 (2002). 

\bibitem{Alain}
A. Barrat and M. Weigt, Eur. Phys. Jour. {\bf 13}, 547 (2000); A. Chatterjee and P. Sen, Phys. Rev. E {\bf 74}, 036109 (2006); D. Watts and S. Strogatz, Nature {\bf 393}, 440 (1998); S. H. Yook, H. Jeong and A. -L. Barabasi, PNAS {\bf 99}, 13382 (2002); M. Barth\'elemy, Europhys. Lett. {\bf 63}, 915 (2003); M Barth\'elemy and A. Flammini, Phys. Rev. Lett. {\bf 100}, 138702 (2008). 


\bibitem{Vespignani2}
V. Colizza, A. Barrat, M. Barth\'elemy and A. Vespignani, PNAS {\bf 103}, 2015 (2006); C. Viboud, et al. Science {\bf 312}, 447 (2006). 
\bibitem{Manna}
S. S. Manna and P. Sen, Phys. Rev. E {\bf 66}, 066114 (2002). 
\bibitem{Newman1}
D. J. Watts, P. S. Dodds and M. E. J. Newman, Science {\bf 296}, 1302 (2002);


\bibitem{Boguna}
M. A. Serrano, D. Krioukov and M. Boguna, Phys. Rev. Lett. {\bf 102}, 058701 (2009). 
\bibitem{Kleinberg}
J. M. Kleinberg, Nature {\bf 406}, 845 (2000). 
\bibitem{Latora}
V. Latora and M. Marchiori, Phys. Rev. Lett. {\bf 87}, 198701 (2001). 

\bibitem{Newman2}
M. T. Gastner and M. E. J. Newman, Eur. Phys. Jour. B {\bf 49}, 247 (2006). 
\bibitem{Havlin2}
G. Li et al. arXiv:0908. 3869 (2009). 
\bibitem{Moreno}
P. Echenique, J. G\'omez-Garde\~nes and Y. Moreno,
Phys. Rev. E {\bf 70}, 056105 (2004).
\bibitem{Motter3}
A. E. Motter and Y.-C. Lai,
Phys. Rev. E {\bf 66}, 065102 (2002).

\bibitem{ising_g}
G. Bianconi, Physics Letters A {\bf 303}, 166 (2002). 
\bibitem{Doro_ising}
S. N. Dorogovtsev, A. V. Goltsev, J. F. F. Mendes, Phys. Rev. E {\bf 66}, 016104 (2002); M. Leone, A. V\'azquez, A. Vespignani and R. Zecchina, Eur. Phys. J. B {\bf 28 }, 191 (2002). 
\bibitem{Doro2}
A. V. Goltsev, S. N. Dorogovtsev and J. F. F. Mendes, Phys. Rev. E {\bf 67}, 026123 (2003). 
\bibitem{Lebellac}
M. Le Bellac, {\em Quantum and Statistical Field Theory} (Oxford University Press, Oxford, 1991); J.W. Negele and H. Orland {\em Quantum many-particle systems} (Addison-Wesley, Reading, MA, 1988).
\bibitem{entropy}
G. Bianconi, EPL, {\bf 81} 28005 (2008); G. Bianconi, Phys. Rev. E {\bf 79}, 036114 (2009);
S. Bradde and G. Bianconi, Jour. Phys. A {\bf 42}, 195007 (2009). 
\bibitem{finite}
S. H. Lee, M. Ha, H. Jeong,  J. D. Noh, H. Park,
Phys. Rev. E {\bf 80}, 051702 (2009).
\end{thebibliography}
\end{document}